%% file: euro_bab.tex
\documentstyle{europhys}
\input euromacr

\begin{document}
\euro{}{}{}{}
\Date{}
\shorttitle{P. GALLO \etal NON-EXPONENTIAL KINETIC BEHAVIOUR OF 
CONFINED WATER}
\title{Non-exponential kinetic behaviour of confined water}
\author{P. Gallo\inst{1}, M. Rovere\inst{1}, M. A. Ricci\inst{1},
     C. Hartnig\inst{2},  E. Spohr\inst{2}}
\institute{
     \inst{1}
          Dipartimento di Fisica, Universit{\`a} di Roma Tre, \\ 
and Istituto Nazionale per la Fisica della Materia, Unit{\`a} di Ricerca
Roma Tre, \\ Via della Vasca Navale 84, I-00146 Roma, Italy \\
     \inst{2} Department of Theoretical
Chemistry, University of Ulm, \\ Albert-Einstein-Allee 11, D-89069 Ulm,
Germany }
%
%%%    The `\maketitle' macro needs the following macro:    \rec{}{}
%%%    to be left empty.
%
\rec{}{}
%
%%%   Physics Abstracts Classification.
%
\pacs{
\Pacs{61}{20Ja}{Computer simulation of liquid structure}
\Pacs{64}{70Pf}{Glass transitions }
\Pacs{61}{25Em}{Molecular liquids} 
      }
\maketitle
%
%%%   The Abstract
%
\begin{abstract}
We present the results of molecular dynamics simulations of SPC/E water 
confined in a realistic model of a silica pore. The single-particle dynamics 
have been studied at ambient temperature for different hydration levels. 
The confinement near the hydrophilic surface makes the dynamic behaviour of 
the liquid strongly dependent on the hydration level. Upon decrease of the 
number of water molecules in the pore we observe the onset of a slow dynamics 
due to the ``cage effect''. The conventional picture of a stochastic 
single-particle diffusion process thus looses its validity.
\end{abstract}
%
%
%%%   Main text
%
  The normal limit of supercooling and the possibility of
  vitrification of water is the subject of a longstanding scientific
  debate~\cite{pablo,angell}. 
  At $T_H\sim236$ K homogeneous nucleation, due to the
  presence of impurities, drives bulk supercooled water into its 
  solid, crystalline phase.
  Experiments sustain the hypothesis 
  that the amorphous phase can be connected to the normal liquid phase
  through a reversible thermodynamic path~\cite{speedy}.  
  Recent molecular dynamics  
  (MD) simulations of bulk supercooled SPC/E~\cite{BGS87} water also showed
  a kinetic glass transition  as predicted by 
  mode coupling theory (MCT)~\cite{goetze}
  at a critical temperature $T_C\sim T_S$~\cite{gallo}, where $T_S$
  is the singular temperature of water~\cite{SpeedyAngell}, 
  which is $T_S=228K$ or $49$ degrees below the temperature of maximum 
  density.

  A comparison of the behaviour of the bulk liquid with the same liquid
  in a confined environment is highly interesting for the development
  of both biological and industrial applications.  Understanding how
  the dynamics of liquid water is perturbed by the interaction with
  hydrophilic or hydrophobic substrates at various levels of hydration
  lays the physical basis for the prediction of stability and
  enzymatic activity of biologically important
  macromolecules~\cite{careri}.  Many experimental studies have been
  performed on confined and interfacial water. They all showed
  evidence of a substantial degree of slowing down of water when
  confined in the proximity of a polar surface~\cite{chen+mc}.
  Particularly significant in this context are the indications of a
  transition of adsorbed water to a glassy state, which is supposedly 
  driven by the protein surface~\cite{doster}.  MD calculations
  found a reduced self diffusion coefficient of water in
  contact with solid hydrophilic surfaces, when 
 compared to bulk water~\cite{rossky}.  Moreover,
  a recent simulation study of water close to the surface of a protein
  evidenced a typical spectral glassy anomaly, the so called {\it
  boson peak}~\cite{bizzarri}, which the authors related to
  protein-solvent coupling.  These and other 
  studies suggest the possibility of a common underlying molecular
  mechanism for the slowing down of the single particle dynamics of
  interfacial water. Nevertheless, many questions concerning the
  substrate-induced perturbation of the dynamics of water are still
  unanswered. Some of these questions can be addressed by MD
  simulation techniques due to their power as a very versatile
  microscopic tool.

  In this letter, we present the results of MD simulations of SPC/E
  water confined in a cylindrical silica pore.  The confining system
  has been carefully designed to mimic the pores of real Vycor glass.
  We chose this particular system, because Vycor is characterized by a
  quite sharp distribution of pore sizes with a small average diameter
  ($\sim 40 \pm5$ \AA), because the pore size does not depend on the
  hydration level (no swelling occurs), and because the surface is
  strongly hydrophilic.  For the same reasons, this system has been
  studied extensively by a variety of
  experiments~\cite{chen+mc,chen,mar1,chenla}.  Here, we focus on the
  role of hydration. In particular, we try to understand the extent to
  which the changes in the system dynamics upon lowering the hydration
  level in a hydrophilic silica pore are equivalent to supercooling
  the bulk phase~\cite{chen,chenla}.  To our knowledge, this is the
  first time that the dynamics of water in a realistic pore model is
  calculated in such detail. The results presented here can help to
  shed light on the microscopic mechanism governing the dynamics of
  water close to a hydrophilic surface.

  The MD simulations started from a silica cube containing a
  cylindrical cavity. In the system preparation step a vitreous silica
  cubic cell was first created by melting a $\beta$-cristobalite
  crystalline structure and then quenching the liquid down to ambient
  temperature~\cite{brodka,feuston}.  Inside the glass, which is
  formed by this process, a cylindrical cavity with a radius
  of $R=20$~{\AA}
  is created by appropriate removal of atoms.  Next, all those
  remaining silicon atoms which are bonded to less than four oxygens
  are removed. With this procedure two kinds of oxygen atoms are
  obtained, {\it nonbridging} oxygens (NBOs), bonded to only one
  silicon ion, and {\it bridging} oxygens (BOs), bonded to at least
  two silicons. The NBOs are saturated with hydrogen atoms, similar to
  the experimental process before hydration~\cite{mar1}, which leads
  to free hydroxyl groups on the pore surface.  With this process, we
  create a cavity with a rough surface, whose geometry, size, and
  microscopic structure are typical for a pore in Vycor glass.  The
  block of silica provides a rigid matrix for the simulation of SPC/E
  water molecules which are then introduced into the pore.  Water
  interacts with the atoms of the rigid matrix by means of an
  empirical potential model~\cite{brodka}, where different fractional
  charges are assigned to BOs ($-0.629 |e|$), NBOs ($-0.533 |e|$),
  silicon ($+1.283 |e|$) and surface hydrogen atoms ($+0.206 |e|$).
  Additional Lennard-Jones potential functions between
  the oxygen sites of water and BOs and NBOs complement the
  water-Vycor interaction~\cite{LJpar}. 
  Since the simulations are time-consuming,
  we have used the shifted force method with a cutoff at $9$~{\AA}.  We
  did check, however, that the use of a larger cutoff or Ewald
  summations does not change the trend of the obtained 
  results~\cite{spohr}.

  There is experimental evidence that the average density of water
  confined in real Vycor glass at full hydration is 
  $11 \%$ less than the density of bulk water at ambient 
  conditions~\cite{c0355}. 
  In order to match this value 2700 water molecules
  are needed in our pore, this number defines our full hydration.  
  We have conducted simulations in the
  NVE ensemble at ambient temperature ($T=298 K$) for four different
  hydrations of the pore.  Simulations with 500, 1000, 1500, 2000
  and 2600 water molecules correspond, based on the above consideration, 
  to hydration levels of 19\%, 37\%, 56\%, 74\% and 96\%, respectively. The
  timestep for the integration of the molecular trajectories is
  $1$~fs. Thermal equilibrium was achieved through the use of a
  Berendsen thermostat~\cite{termo}; the equilibration process has
  been carefully monitored via the time dependence of the potential
  energy.  Further details of the simulation can be found in
  ref.~\cite{spohr}, where static properties of the system are
  studied, and where it is shown that the interaction model used here
  leads to satisfactory agreement with the experimental~\cite{mar1}
  site-site pair correlation functions.

  We now discuss some of the most significant single particle
  dynamical properties of water
  from our simulations for different hydration levels at ambient
  temperature.  In Fig.~\ref{fig:R2} we show the mean square
  displacement (MSD) of the center of mass motion along the
  non-confined z-direction.  
  After an initial ballistic diffusion, water molecules in the bulk
  phase at ambient temperature (not shown) enter the Brownian
  diffusive regime (characterized by a slope of 1 in the log-log
  plot).  Our data show at intermediate times an obvious flattening 
  of the MSD curve, which increases with decreasing hydration level.  
  This is the signature of the so-called cage effect in the system, 
  which describes the trapping of molecules in the cage formed by 
  their nearest neighbors at intermediate times after the initial 
  ballistic regime. 
  Once the cage has relaxed, molecules enter the normal diffusive
  regime.  In the inset of Fig.~\ref{fig:R2} we report the self
  diffusion coefficient, $D$, as calculated from the slope of the MSD
  in the diffusive regime. A substantial decrease of $D$ with
  decreasing hydration level is evident. All values of $D$ are
  substantially lower than the one of SPC/E bulk water at the same
  temperature. 

  Figure~\ref{fig:FQT} shows the single particle intermediate
  scattering function (ISF), $F_S(Q,t)$, calculated for the center of
  mass motion at the maximum of the oxygen-oxygen structure factor for
  all hydration levels investigated. As the hydration is lowered, the
  ISF displays an increasingly pronounced shoulder around 1 ps, which
  is the characteristic of a two-step relaxation process.  The long
  time tail is highly non-exponential for all hydration levels \cite{NGP}. 
  We are able to fit a combination of the Kohlrausch-William-Watts (KWW) 
  relaxation function plus a gaussian term 
  \begin{equation}
  F_S(Q,t)=
  \left[ 1-A(Q) \right] \exp \left[ {-\left( t/\tau_s \right)^2} \right]+
  A(Q) \exp \left[ {- \left( t/\tau_l \right)^\beta} \right]
  \label{strexp}
  \end{equation}
 only to the ISF at the lowest hydration level. 
 $A(Q)=e^{-a^2q^2/3}$ is the Debye-Waller factor arising from the
 cage effect. $a$ is
 the amplitude of the vibration of the molecule
inside the cage created by the potential barrier of the nearest neighbours.
 $\tau_s$ and $\tau_l$ are, respectively, the short and
 the long relaxation times.
 The fitted curve is shown as the dashed line in 
 Fig.~\ref{fig:FQT}.  For bulk water, this function fitted the center
of mass ISF for several temperatures \cite{gallo}.  For the
shown fit we obtain a cage radius $a\simeq0.44$~{\AA},
which is similar
to the radius obtained for bulk supercooled water, $a\simeq0.5$~{\AA}.
The lower radius obtained for confined water may be 
due to the slightly higher density of water close to the surface
\cite{spohr}.  The short relaxation time $\tau_s\simeq 0.14$~ps is
again comparable to the bulk value $\tau_s\simeq 0.2$~ps.  
The long time tail is characterized by $\beta=0.35$ and  $\tau_l=356$~$ps$. 

For the higher hydration levels
we have calculated the ISF separately 
for the first two layers of molecules
close to the surface and for the remaining molecules.
This choice is based on the shape of the density profile, which 
shows a double layer of water molecules close to the surface
with density higher than the average \cite{spohr}. This double layer
extends up to $5$~{\AA} from the surface.
The inset of Fig.~\ref{fig:FQT} shows, for $N_W=2600$  (96\% hydration), 
the two contributions. While the molecules in the adsorbed layer
relax very slowly, the decay of the ISF of the remaining molecules
is much faster and very well described by eq.\ref{strexp} 
(bottom curves).
From the fit we extract $\beta=0.71$, $\tau_l=0.5$~$ps$, $\tau_s=0.17$~ps,
and $a=0.54$~{\AA}. While the $a$, $\tau_s$ and $\tau_l$ are similar to 
the values found for bulk water at ambient conditions, the
$\beta$ differs from the value, $1$, of the bulk.
It turns out that the behaviour of molecules belonging to the first
hydration layers change as a complete surface coverage is achieved.
In fact for $N_W=500$ the surface coverage is not complete and 
patches of water molecules are visible along the pore surface. 
A stretched exponential function is able to account also for
the dynamics of clusters in a frozen environment \cite{goetze}.
In Fig.~\ref{fig:BP} $F_S(Q,t)$ at 19 \% hydration is shown for several
values of $Q$ ranging from 0.5 to 4 ${\rm\AA}^{-1}$.  This figure very
clearly demonstrates the overshooting of the ISF around 1 ps, which
was also visible in Fig.~\ref{fig:FQT}.  

The maximum can be observed
both in the $z$ direction along the pore axis and in the $xy$ plane,
as is visible in the inset.  The intermediate maximum of the ISF has
been related to the so-called boson peak (BP) \cite{binder}.  The BP
is an excess of vibrational modes present in
many glasses at frequencies around $1$~THz.  When this 
glassy anomaly appears in
a liquid phase, it is usually considered as a precursor to the actual
glass transition.  
We also performed a shell analysis of the
dynamic behaviour at the higher hydration levels, and we found 
in these cases that the contribution to the BP in our simulation 
comes only from water molecules which are not in the first layer.  
This fact, and the fact that our
substrate is a rigid framework, is an indication that the BP is a
feature of liquid water, which is not induced by the substrate
dynamics.

Experimental signatures of our observations were detected in some
confined hydrogen bonded complex liquids~\cite{melni}.  
An analysis of quasi-elastic neutron
scattering data of the water~/~Vycor system is consistent with a highly
non-exponential relaxation behaviour \cite{chenla}.
The BP for the water~/~Vycor system has been 
detected recently at energies around $3.5$~meV~\cite{canni}.  

In summary, we presented MD results concerning the 
single particle dynamics of liquid water confined in a silica pore.
We found evidence of glassy behaviour already  at room temperature.
On lowering the hydration level of water inside the pore, 
the MSD flattens at intermediate times due to a cage effect. 
Correspondingly, the ISF displays a two step relaxation behaviour 
with a highly non-exponential slow relaxation.
Such behaviour is typical for a glass forming liquid
approaching its glass transition point.
At the lowest hydration level, a KWW function could be fitted to the 
ISF. In this particular system, all phenomena are  strongly influenced 
by the substrate. Specifically the interaction 
with the hydrophilic surface seems to drive the liquid closer to the
glass transition point.  This interaction leads to a significant
slowing down of water molecules close to the substrate. 
The strong distortion of the hydrogen bond network close to the
surface will lead to a strong variation of the cage structure from one
molecule to the other, which, in contrast to bulk water, might mask
the simple behaviour suggested by mode coupling theory.
The dynamic results moreover support the notion of two quite distinct
subsets of molecules. One is bound directly to the substrate surface,
the other consists of the remaining water molecules in the pore
center.  Near the substrate surface, the local water density is
slightly higher \cite{spohr}, and the dynamics is severely slowed down
at all hydration levels, whereas the retardation of the slow
relaxation process is much less pronounced for the inner
water shells, whose contribution to the slow relaxation in the ISF
could be fitted to a KWW law. 
The shouldering of the relaxation law is indeed 
determined by the propagation of the perturbation of the substrate
to the water molecules 
inside the pore. With the increase of the hydration level the effect
is overthrown by the increasing number of molecules located
far from the surface.

The confinement also leads to the formation of an 
overshooting related to the BP at the
lowest hydration level. Since the BP is found typically in
strong glass formers \cite{binder}, the appearance of the intermediate
maximum in the ISF at the lowest hydration would favor the hypothesis
of a conversion to a more strong glass-forming behaviour of water as the 
level of hydration is lowered.

\stars
P.G. wishes to thank S.-H. Chen for all the interesting discussions
on this subject.
%
%%%   Bibliography environment begins here. You can use the macros \Name{},
%%%   \And, \Book{} or \Review{}, \Vol{}, \Year{} and \Page{}, to type your
%%%   references.
%
\newpage
\begin{figure}
%
%\vbox to 1cm{\vfill\centerline{\fbox{Here is the figure 1}}\vfill}
%
\caption{Log-log representation of the mean square displacement as 
a function of time 
for the hydration levels studied. From bottom to top the curves 
correspond to increasing
 hydration levels. The inset shows self diffusion coefficients, $D$, vs. 
number of molecules in the pore (filled circles).
The line is a guide for the eye, and the filled diamond on the y-axis
is the value of $D$ for SPC/E bulk water at ambient conditions.
}
\label{fig:R2}

%
%\vbox to 1cm{\vfill\centerline{\fbox{Here is the figure 2}}\vfill}
%
\caption{
Intermediate scattering function (ISF) of the center of mass motion at $Q=2.25\ {\rm \AA}^{-1}$,
corresponding to the peak of the structure factor.
The ISF is shown over four decades; full lines correspond, from top to bottom,
to increasing hydration levels. 
The dashed line is the fit of 
Eq. \protect\ref{strexp} 
to the data at 19 \% hydration. 
From the fit we extract $\beta=0.35$, $\tau_l=356$~$ps$, $\tau_s=0.14$~ps,
and $a= 0.44 $~{\AA} (see also text).
The inset shows the ISF at 96 \% hydration. 
The top curve 
(labeled S)
is the contribution to the total ISF  from the 
shell of molecules closest to the substrate ($15<r<20$~{\AA}), the bottom curve 
(labeled C) 
the contribution from the molecules in the remaining shells
($0<r<15$~\AA), 
and the central one 
(labeled T) is the total ISF. 
The radius of the pore is $R=20$~{\AA} and the zero is taken 
at the center of the pore.
The dashed line is the fit of 
Eq. \protect\ref{strexp}. 
From the fit we extract $\beta=0.71$, $\tau_l=0.5$~$ps$, $\tau_s=0.17$~ps,
and $a=0.54$~\AA .
}
\label{fig:FQT}
%
%\vbox to 1cm{\vfill\centerline{\fbox{Here is the figure 3}}\vfill}
%
\caption{
Self contribution to the intermediate scattering function (ISF) at
19 \% hydration for various values of the scattering vector $Q$
in the z direction. From top to bottom, 
$Q=0.5,1,1.2,1.5,1.8,2,2.25,2.5,3,3.5,4$~$\rm\AA^{-1}$. 
The inset shows the ISF for the components of $Q$ 
along the $z$ direction (full line), in the $xy$ plane (dotted-dashed
line), and the sum of the two (dashed line) for two $Q$ values.
The first three curves on the top correspond to  
$Q=2.25$~$\rm\AA^{-1}$, i.e. the peak of the structure factor,
and the three bottom ones to $Q=4$~$\rm\AA^{-1}$. 
The maximum related to the boson peak (BP) is evident at around
$1$~ps.
}
\label{fig:BP}
\end{figure}

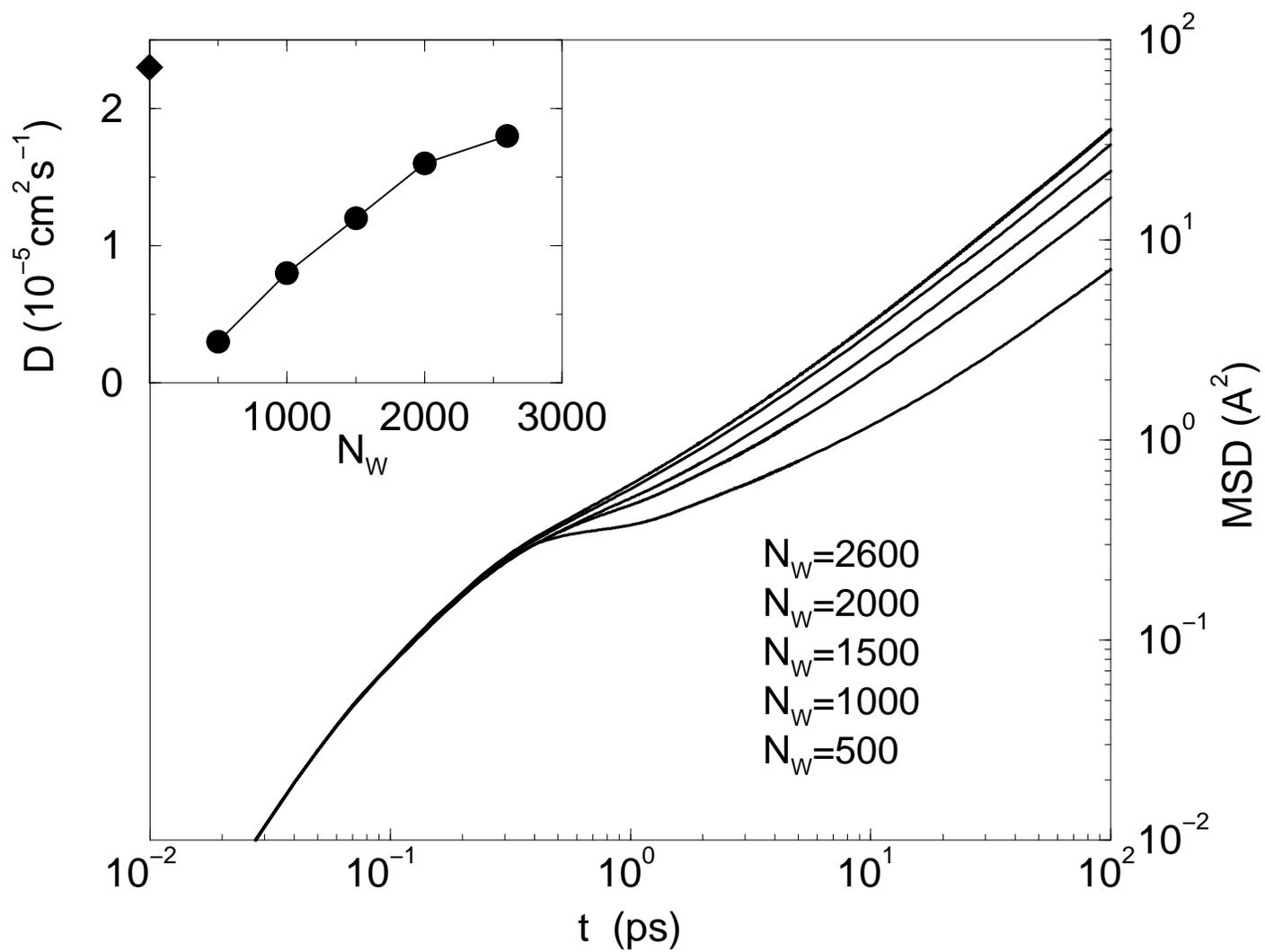
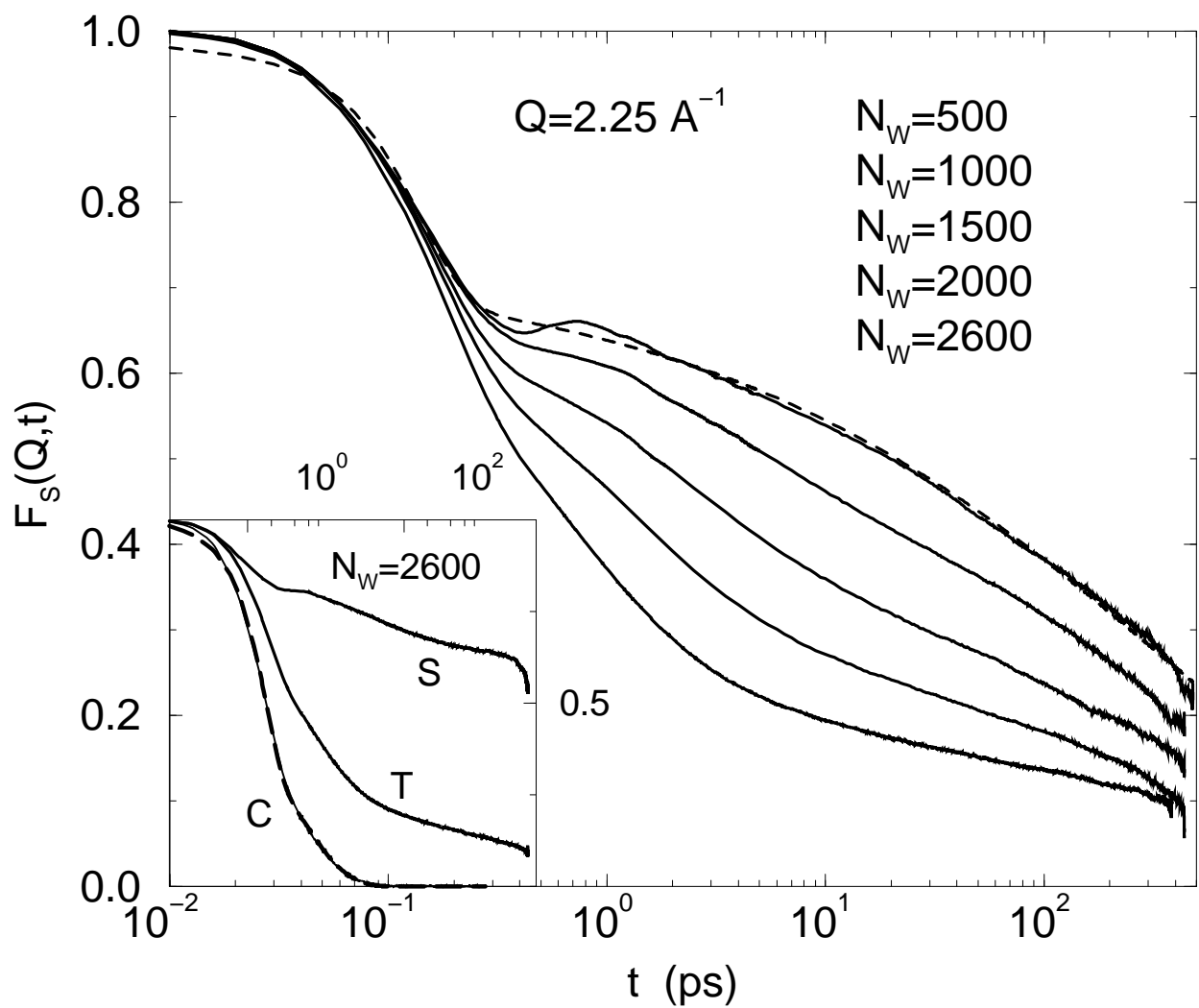
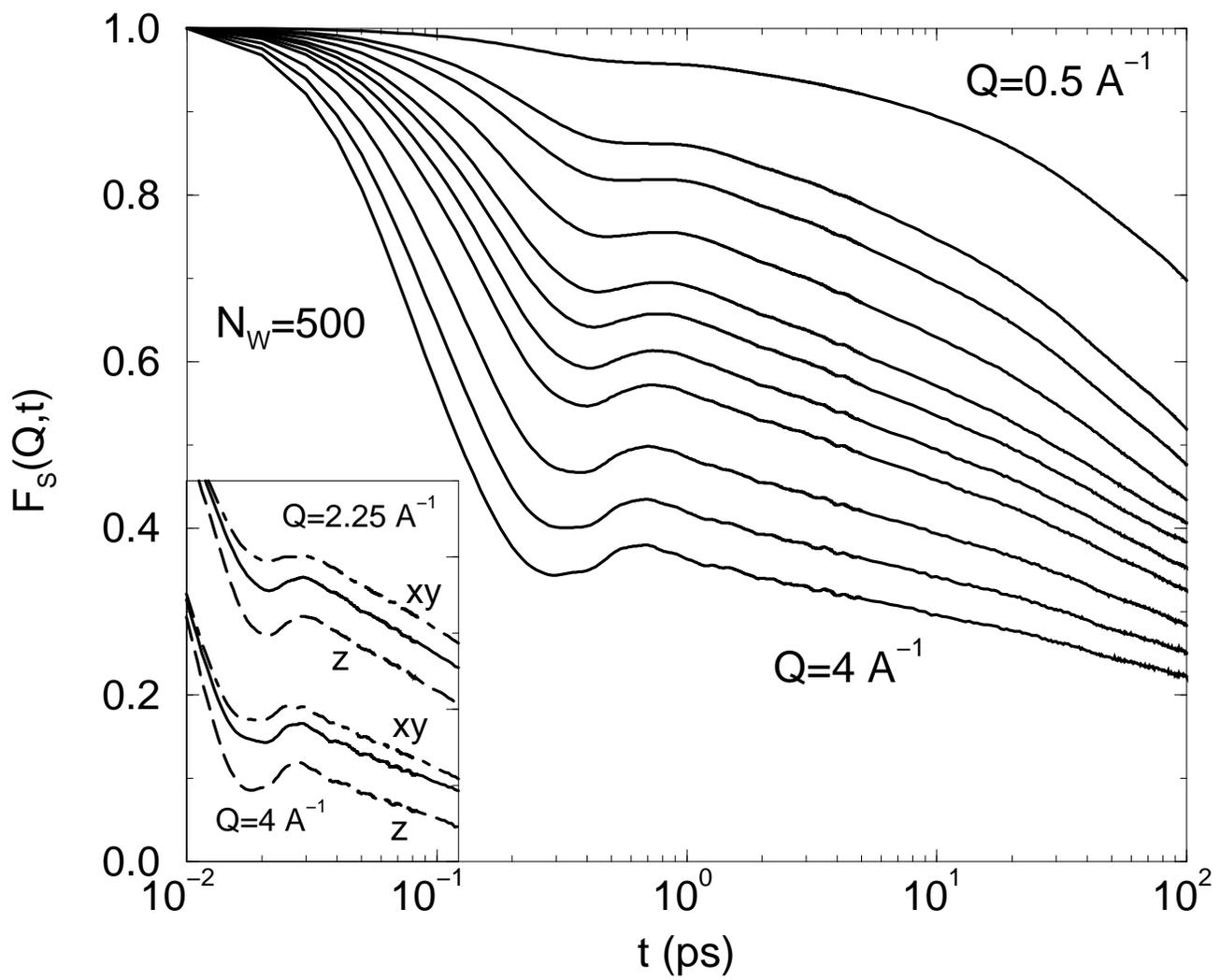
\end{document}

%% file: euromacr.tex
\def\etal{{\hbox{{\tenit\ et al.\/}\tenrm :\ }}}

\def\And{{\rm and\ }}

\def\stars{\bigskip\centerline{***}\medskip}

\newif\ifboo \boofalse

\def\Review#1{\boofalse{\it #1},}
\def\Name#1{{\sc #1},}
\def\Vol#1{\ifboo Vol. {\bf #1}\else{\bf #1}\fi}
\def\Year#1{\ifboo #1\else(#1)\fi}
\def\Book#1{\bootrue{\it #1},}
\def\Page#1{\ifboo {\rm p. #1}\else{\rm #1}\fi}